\documentstyle[12pt]{article}

\parskip=\medskipamount            
\def\7#1#2{\mathop{\null#2}\limits^{#1}}        

\def\beee{\begin{equation}}
\def\eeee{\end{equation}}
\def\dggg{^{\dagger}}
\begin{document}

\bibliographystyle{unsrt}

\begin{center}

{\bf QUON STATISTICS FOR COMPOSITE SYSTEMS AND A\\
LIMIT ON THE VIOLATION OF THE PAULI PRINCIPLE FOR\\ NUCLEONS AND
QUARKS}\\[.1in]

O.W. Greenberg\footnote{email address, owgreen@physics.umd.edu.}\\
Center for Theoretical Physics\\
Department of Physics \\
University of Maryland\\
College Park, MD~~20742-4111\\

and\\

Robert C. Hilborn\footnote{email address, rchilborn@amherst.edu}\\
Department of Physics\\
Amherst College\\
Amherst, MA 01002-5000\\[5mm]

University of Maryland Physics Paper 99-089\\

quant-ph/9903020

\end{center}

\vspace{5mm}

\begin{abstract}

The quon algebra gives a description of particles, ``quons,'' that are
neither fermions nor bosons.  The parameter $q$ attached to a quon
labels a
smooth interpolation between bosons, for which $q = +1$, and fermions,
for
which $q = -1$.  Wigner and Ehrenfest and Oppenheimer showed that a
composite system of identical bosons and fermions is a fermion if it
contains an odd number of fermions and is a boson otherwise.  Here we
generalize this result to composite
systems of identical quons.  We find
$q_{composite}=q_{constituent}^{n^2}$
for a system of $n$ identical quons.
This result reduces to the earlier result for bosons and fermions.
Using
this generalization we find bounds on possible violations of the Pauli
exclusion principle for nucleons and quarks based on such bounds for
nuclei.

\end{abstract}

The celebrated spin-statistics
theorem \cite{sst} gives the spin-statistics connection that
lies at the heart of the quantum mechanics of
many-body systems:  spin one-half particles are fermions and integer
spin particles
are bosons.  Wigner \cite{w} and Ehrenfest and Oppenheimer \cite{eo}
found the fundamental result
that composite systems of identical bosons and fermions are bosons,
unless they
have an odd number of fermions, in which case they are fermions.  Given
the usual rules for addition of angular momenta, the
Wigner--Oppenheimer-Ehrenfest (WEO) result extends the spin-statistics
connection to composite systems. The WEO result requires
that all the composites must be in the same internal quantum state and
that the
interactions between the composites must be sufficiently weak not to
change the internal states of the composites.  In this letter we
generalize the WEO result to cases where the bound states are composed
of
identical particles that need not be either bosons or fermions.

Recently
one of us introduced a formalism, using the quon algebra, that can
describe small
violations of the usual spin-statistics connection \cite{g}.  In this
formalism
identical particles can occur in states associated with a mixture of
different representations of the
symmetric group, with the weights of the mixture dependent on a
parameter $q$.
To introduce the quon algebra, recall that
bosons are described by the algebra of commutators of the creation and
annihilation operators
\beee
a(k) a\dggg(l) -a\dggg(l) a(k) = \delta(k,l),       \label{comm}
\eeee
and fermions are described by the algebra of anticommutators of the
creation and
annihilation operators,
\beee
a(k) a\dggg(l) +a\dggg(l) a(k) = \delta(k,l).       \label{ac}
\eeee
The quon algebra is the convex sum of these
two expressions, where Eq.(\ref{comm}) is multiplied by $(1+q)/2$,
Eq.(\ref{ac})
is multiplied by $(1-q)/2$, and the equations are added.  For the sum to

be
convex $q$ must be between $-1$ and $1$.  Several authors have shown
that when
$q$ is in this range all the squared norms of states are positive.
\cite{pos}
Thus the quon algebra is
\beee
a(k) a\dggg(l) -q a\dggg(l) a(k) = \delta(k,l).       \label{q}
\eeee
Just as in the case of Bose or Fermi statistics we choose the Fock-like
representation in which
\beee
a(k)|0 \rangle =0.
\eeee
For $q=1$ we recover Bose statistics
and for $q=-1$ we get Fermi statistics.
The probabilities for two identical particles that are quons
to be in the symmetric or the antisymmetric state can be found
by calculating the following matrix element using the rules just given,
\beee
(a\dggg (k_1) a\dggg(k_2) |0\rangle,a\dggg (l_1) a\dggg(l_2) |0\rangle)
= \nonumber \\
\eeee
\beee
\delta(k_1,l_1)\delta(k_2,l_2)+q \delta(k_1,l_2)\delta(k_2,l_1) =
\nonumber \\
\eeee
\beee
\frac{1+q}{2}
[\delta(k_1,l_1)\delta(k_2,l_2)+\delta(k_1,l_2)\delta(k_2,l_1)]
+\frac{1-q}{2}
[\delta(k_1,l_1)\delta(k_2,l_2)-\delta(k_1,l_2)\delta(k_2,l_1)],
\label{2}
\eeee
where we recognize the quantities in the square brackets as the
scalar products of the
symmetric and antisymmetric states of two particles.  For systems with
more than two identical particles there will also be many-dimensional
representations of the symmetric group.  These occur with $q$-dependent
probabilities.  For $q \rightarrow 1 ~~(-1)$ the more symmetric
(antisymmetric)
representations are more heavily weighted.  To simplify
notation we assume that spin and/or isospin variables and
other variables are included in the space or momentum variables.

We want to find a result for the quon statistics of
composite states analogous to that of
Wigner and of Ehrenfest and Oppenheimer under the corresponding
conditions
for the internal states and their interactions.  First, we note that in
generating the Fock-like states from the vacuum the order of operators
is
significant in a composite state of quons;
in fact for an $n$-body composite all $n!$ states with permutations
of the orders of the creation operators are linearly independent
(provided all the operators carry different quantum numbers).  This
occurs
because quons can occupy states associated with all the representations
of
the symmetric group.
Secondly, we point out that a bound state of quons will be in a
single irreducible representation, $r$, of the symmetric group and thus
its wave function will have a linear combination of terms with the
creation operators in permuted orders.  Since matter is made of
electrons
and nucleons which are both fermions or, if quons, very close to
fermions,
the most relevant case is when the quons are in an antisymmetric state,
$r=a$,
i.e., have  $q$ close to $-1$.  Our purpose here is to find the
statistical
behavior
of many-particle systems of such bound states.  To our surprise,
however, the
statistical behavior of the many-quon bound states is independent of
$r$.
The only way the representation $r$ enters is in the normalization of
the
individual bound states.  To make this clear we first consider a single
bound state.

For a nonrelativistic
theory we can represent the bound-state creation operator in terms
of the creation operators for the constituents as
\beee
b\dggg (x)=N_r\int \prod_{i=1}^n d^3y_i
f^{(n)}(x-y_1, \cdots, x-y_n) \sum_P c_r(P):a\dggg(y_{P1}) \cdots
a\dggg(y_{Pn}):
\eeee
or, in momentum space, as
\beee
\tilde{b}\dggg (p)=\tilde{N}_r\int \prod_{i=1}^n d^3k_i
\delta(p-\sum_1^n k_i)
\tilde{f}^{(n)}(k_1, \cdots, k_n) \sum_P c_r(P):\tilde{a}\dggg(k_{P1})
\cdots
\tilde{a}\dggg(k_{Pn}):.
\eeee
 The sums run over the $n!$ permutations of the symmetric group $S_n$
and
our permutations are
place permutations rather than label permutations, since place
permutations
are defined even when some of the quantum numbers carried by the
operators
are the
same.  The coefficients $c_r(P)$ pick a sum of products of operators
that are
in the representation $r$ of $S_n$.

 To normalize the bound state we require
\beee
(b\dggg (x)|0\rangle, b\dggg(x^{\prime})|0\rangle=\delta(x-x^{\prime}).
\label{norm}
\eeee
To avoid repetition we only discuss the situation in $x$-space; the
calculations in
momentum space are similar.  The normalization condition has the form
\beee
|N_r(q)|^2 {\cal P}_r(q)=1,      \label{normal}
\eeee
where ${\cal P}_r$ depends on $n$ (it has degree $n(n-1)/2$ in $q$)
and also depends on $f^{(n)}$.
The $q$-dependence comes from the scalar product
\beee
(:a\dggg(y_{P1}) \cdots a\dggg(y_{Pn}):|0\rangle,
:a\dggg(y_{P1^{\prime}}) \cdots a\dggg(y_{Pn^{\prime}}):|0\rangle).
\label{sp}
\eeee
The graphical rules given in \cite{g} allow calculation of this scalar
product.
Place the operators on one
side of the scalar product on one line and those on the other side
of the product on a similar line below the first.  Each contribution
to the vacuum matrix element comes from a product of contractions
in which a creation operator on the top line is paired with a
creation operator on the line below to give a $\delta$-function.
Each set of contractions is associated with a permutation $R$ that takes

the operators in their original order on the top line and rearranges
them to the order of the operators with which they are contracted.
The power of $q$ associated
with the product of $\delta$-functions coming from a given set of
contractions is the inversion number $i(R)$ of this permutation,
which equals
the minimum number of crossings of these lines.  Since
each operator on one line can pair with every operator in the other
line,
there will be $n!$ terms.  This rule is a generalization of Wick's
theorem in which the usual plus or minus signs are replaced by
$q^{i(R)}$.

We now calculate the analog of Eq.(\ref{2}) with the composite
$b\dggg$'s replacing the $a\dggg$'s.  The essential issue in the
calculation concerns the algebra of the creation operators; the
wavefunctions
just play the role of deciding which operators are connected with a
given
bound state.  Thus only the operator part of the calculation is
relevant.  There will be $(n!)^2$ terms in which the order
of the bound states on the two sides of the scalar product
is preserved but
the order of the operators in each of the two bound states
is permuted in
$n!$ ways.  These terms will give two factors of the polynomial
${\cal P}_r$ that we found in the normalization of the bound
state, one for each bound state.  The normalization factors
obey Eq.(\ref{normal}) so they exactly cancel ${\cal P}_r$.

The contributions to the scalar product when the bound
states are interchanged have a relative factor
of $q^{n^2}$ in addition to the factor of ${\cal P}_r^2$ just
discussed.  To see this consider the case in which the
bound states are interchanged, but the order of the operators
inside each bound state is preserved.  Draw lines from each operator
in the left bound state on the top line to each operator in the
right bound state on the bottom line preserving the order of the
operators in the bound states.  Do the same from each operator in the
right bound state on the top line to the left bound state on the
bottom line.  The $n^2$ crossings give the relative factor of
$q^{n^2}$ between the terms that are interchanged and those that are
not.
A similar argument shows that there is a relative factor of $q^{n^2}$
for all the terms.

The remaining $(2n)!-2(n!)^2$ terms correspond to cases
in which creation operators in one bound state on one
side of the scalar product contract with creation operators in
{\it both} bound states on the other side of the scalar product.
These terms don't correspond to interchange of the bound state just
as they do not in the usual Bose and Fermi cases.  Here and below we
can drop these
terms provided we assume, as in the usual case for the statistics of
bound states, that the interaction between the bound states is
negligible compared to the interaction that binds the constituents
in each bound state.  Thus we have
established that the analog of Eq.(\ref{2}) for the bound states is
\beee
(b\dggg(p_1) b\dggg(p_2) |0\rangle,b\dggg (r_1) b\dggg(r_2)
|0\rangle)
= \nonumber \\
\eeee
\beee
\delta(p_1,r_1)\delta(p_2,r_2)+q^{n^2} \delta(p_1,r_2)\delta(p_2,r_1)
 = \nonumber \\
\eeee
\beee
\frac{1+q^{n^2}}{2}
[\delta(p_1,r_1)\delta(p_2,r_2)+\delta(p_1,r_2)\delta(p_2,r_1)]
+\frac{1-q^{n^2}}{2}
[\delta(p_1,r_1)\delta(p_2,r_2)-\delta(p_1,r_2)\delta(p_2,r_1)].
\label{c}
\eeee

 We can now represent the $n!$ lines that connect a bound state on the
left
hand side of a scalar product to a bound state on the right hand side by
a
superline.  When we use our generalized Wick's theorem to calculate
matrix
elements of the bound state operators we find the same result as for a
single
quon operator except that the number of crossings of superlines replaces
the
number of crossings of lines so that
$q^{n^2}$ replaces $q$.   Thus
bound states of $n$ quons with parameter $q$ have quon statistics with
parameter $q^{n^2}$. \cite{gns,poly}  This generalization of the
Wigner--Ehrenfest-Oppenheimer rule is the main result of this letter.

 Since Bose and Fermi statistics correspond to $q=1$ and $q=-1$,
respectively, we expect the Wigner--Ehrenfest-Oppenheimer rule
to emerge in the limit $q\rightarrow \pm 1$.  Indeed this is true
since $n$ and $n^2$ are even or odd together.

The composite statistics rule is important
in establishing the statistical behavior of identical nuclei in
molecules for example.
Similar issues arise in the collective behavior of Cooper pairs in
BCS-type
superconductivity \cite{schrieffer} and superfluid behavior in $^{3}He$
\cite{who?}.
The crucial
condition for this rule, as pointed out by Ehrenfest and Oppenheimer, is

that all the
composites must be in the same internal state and the interactions
between the
composites must be sufficiently weak so as not to change the internal
states.  In other
words the interaction energies among the composites must be small
compared to the
energy separation of the internal states.  This condition is obviously
well satisfied for
nuclei in molecules where the interactions are on the order of electron
volts and the
internal energy differences are on the order of MeV.  On the other hand
nitrogen
atoms taken individually are fermions (seven electrons, seven protons
and seven
neutrons for $^{14}N$) but the $N_2$ molecule is permutation symmetric
under
the interchange
of the two $^{14}N$ nuclei; the statistics of the individual atoms is no

longer relevant once
the atoms interact to form the molecule.

 We now apply the composite statistics rule to nuclei.  We don't expect
experiments using bound states of many
electrons, i.e., atoms, to improve the extremely high-precision
bound for electrons due to E. Ramberg and G.A. Snow \cite{rs}.
On the other hand experiments with nuclei may give significant bounds on
the
statistics of nucleons, for which there are at present no direct
precision
experimental tests.  For nuclei that are
bound states of $A$ almost-Fermi nucleons the result
is immediate.  If $A$ is even so
the nucleus is close to Bose with parameter $1 \geq q_B=1-\epsilon$,
then the constituent nucleons are quons with parameter
$-1 \leq q_{nucleon}=-1+\epsilon/A^2$.  If $A$ is odd so
the nucleus is close to Fermi with parameter $-1 \leq q_F=-1+\epsilon$,
then the constituent nucleons are quons with parameter
$-1 \leq q_{nucleon}=-1+\epsilon/A^2$.  Thus the bound on the deviation
of the nucleons from Fermi statistics is {\it improved} over the bound
on the deviation of the nucleus by a factor of $1/A^2$.  This analysis
can
be extended to quarks bound in nucleons or other baryons, for which
$q_{quark}^9=q_{baryon}$.

 Empirical bounds on violations of statistics found up
to 1989 are reviewed in
\cite{owg-rnm}.  More recent experiments have set stringent upper
limits on the violation of
the spin-statistics connection for electrons \cite{rs,d} and $^{16}O$
nuclei
\cite{angelis,hilborn}.  DeMille and Derr have found a bound on
violation
of Bose statistics for photons \cite{ded}.  From a sensitive $C^{16}O_2$

molecular spectroscopy experiment Modugno, Ingusicio, and Tino
\cite{modugno} have
been able to show that the probability of finding the two $^{16}O$
nuclei
(spin 0) in carbon
dioxide in a permutation antisymmetric state is less than $5 \times
10^{-9}$.

The $^{16}O$ nucleus is of course a composite system of eight protons
and
eight neutrons.  Applying the results derived above, we can interpret
the
experimental limit
on spin-statistics violations for $^{16}O$ as setting a limit of {$1 +
q_{nucleon}\leq 2 \times 10^{-11}$} for nucleons.  We believe this is
the first
precision limit on the violation of the spin-statistics connection for
nucleons linked directly to an experimental measurement.  Other limits
on
violations of the Exclusion Principle for nucleons have relied on
arguments
based on models of the solar p-p cycle \cite{Plaga} (from which a limit
of
about $10^{-15}$ is claimed) or arguments based on models for isotopic
elemental
production in supernovae core collapses \cite{baron}.  In the latter
case
there are many uncertainties in the model for the collapse and the
precision mass spectrometry searches for unusual isotopes have yet to be

done.  Extending the analysis to quarks within the model of nucleons as
bound states of 3 quarks, yields $1 + q_{quark}\leq 2 \times 10^{-12}$
for
quarks.

There is yet another way to extend existing experimental limits.  In a
recent paper Greenberg and Hilborn \cite{grh} have
argued
that an experimental limit on the spin-statistics violations for one
type of particle, say
electrons, can be used to set a limit on spin-statistics violations for
particles with which
the original particles interact.  For example it is possible to extend
the stringent limit $1+q < 10^{-26}$ for a spin-statistics violation for

electrons from the Ramberg-Snow
experiment \cite{rs} to set a limit on a possible spin-statistics
connection violation for photons.

The theorem proved in this paper provides the quon generalization of the

standard rule for the statistics of composite systems (fermion behavior
for an odd
number of fermions; otherwise boson behavior occurs).
In accord with the original
work by Wigner and by Ehrenfest and Oppenheimer, we have shown that the
generalized rule $q_{composite}=q_{constituent}^{n^2}$ holds as
long as the composites are all in the same internal state and the
interaction energies
among the composites can be ignored compared to the differences in
internal state energies (i.e. the composites are tightly bound).


\begin{thebibliography}{99}
\bibitem{poly} To impose the condition that the quons are in an
antisymmetric
(symmetric) state,
we simply choose the $f^{(n)}$ to be antisymmetric (symmetric).  Then
the $q$-dependent terms factor from the terms that depend
on the wave function. Explicit forms for
the analog of ${\cal P}_a$ (or, ${\cal P}_s$) for the first few values
of
$n$ are $1$, $1-q$, $1-2q+2q^2-q^3$. (For the
symmetric case, replace the $-$ signs by $+$ signs.)
\end{thebibliography}
\end{document}